\documentclass[12pt,a4]{article}
\usepackage{setspace}
\usepackage{amssymb}
\usepackage{amsmath,amssymb}
\usepackage[T1]{fontenc}
\usepackage[bf,small,tableposition=top]{caption}
\usepackage{subfig}
\usepackage{amsthm}
\usepackage{lineno}
\usepackage[figuresright]{rotating}
\usepackage{enumerate}
\usepackage{authblk}
\usepackage[T1]{fontenc}
\usepackage[utf8]{inputenc}
\usepackage{graphics}
\usepackage{graphicx}
\usepackage{multirow}
\usepackage{lscape}
\usepackage[dvipsnames]{xcolor}
\usepackage{mathrsfs}
\usepackage{hyperref}
\usepackage{textcomp}
\usepackage{listings}
\usepackage{multirow}

\begin{document}
\title{Analysis of Count Data by Transmuted Geometric Distribution}

\author[1]{Subrata Chakraborty}
\author[2]{Deepesh Bhati\footnote{deepesh.bhati@curaj.ac.in(Corresponding Author)}}
\affil[1] {\textit{Department of Statistics, Dibrugarh University, Assam, India}}
\affil[2] {\textit{Department of Statistics, Central University of Rajasthan}}
\maketitle

\begin{abstract}
Transmuted geometric distribution ($\mathcal{TGD}$) was recently introduced and investigated by Chakraborty and Bhati (2016). This is a flexible extension of geometric distribution having an additional parameter that determines its zero inflation as well as the tail length. In the present article we further study this distribution for some of its reliability, stochastic ordering and parameter estimation properties. In parameter estimation among others we discuss an EM algorithm and the performance of estimators is evaluated through extensive simulation. For assessing the statistical significance of additional parameter$(\alpha)$, Likelihood ratio test, the Rao's score tests and the Wald's test are developed and its empirical power via simulation were compared. We have demonstrate two applications of ($\mathcal{TGD}$) in modeling real life count data.
\end{abstract}


\section*{Introduction}
Chakraborty and Bhati (2016) recently introduced the transmuted geometric distribution $\mathcal{TGD}(q,\alpha)$ using the quadratic rank transmutation techniques of Shaw and Buckley (2007). It may be noted that though there is a large number of new continuous distribution in statistical literature which are derived using the rank transmutation technique but $\mathcal{TGD}(q,\alpha)$ is the first discrete distribution derived using this technique.
Chakraborty and Bhati (2016) investigated various distributional properties, showed applicability of $\mathcal{TGD}(q,\alpha)$ in modeling aggregate loss, claim frequency data from automobile insurance and demonstrated the feasibility of $\mathcal{TGD}(q,\alpha)$ as count regression model by considering data from health sector. As $\mathcal{TGD}(q,\alpha)$ is a simple yet elegant extension of the celebrated geometric distribution with potential of application in various context of discrete data analysis. In the current article, we discussed some additional theoretical and applied aspects of $\mathcal{TGD}(q,\alpha)$, which are structured as follows. In section 2, we present various reliability properties and stochastic ordering of $\mathcal{TGD}(q,\alpha)$. In section 3, comparative study of maximum likelihood estimator(ML) obtained numerically and through EM Algorithm are presented through simulation, whereas in section 4, detailed hypothesis testing is discussed considering three Wald's, Rao's Score and Likelihood Ratio test for testing $\alpha=0$. To illustrate the applicability of $(\mathcal{TGD})$ models in different disciplines other than those discussed in Chakraborty and Bhati (2016), we consider two real data sets and compare them with different family of distributions in Section 5. Finally, some conclusions and comments are presented in Section 6.

\section{Transmuted geometric distribution ($\mathcal{TGD}(q,\alpha)$)}
\noindent A random variable (rv) $X$ is said to follow \textbf{Transmuted geometric distribution} $(\mathcal{TGD})$ with two parameters $q$ and $\alpha$, in short, $\mathcal{TGD}(q,\alpha)$ if its probability mass function (PMF) is given by
\begin{equation} \label{e2}
p_y=\left(1-\alpha\right)q^y(1-q)+\alpha(1-q^2)q^{2y}, \quad y=0,1,\cdots .
\end{equation}
The corresponding survival function (sf) is written as
\begin{equation} \label{sf}
\bar{F}_Y(y)=(1-\alpha)q^{y}+\alpha q^{2 y}, \quad y=0,1,\cdots .
\end{equation}
where $0<q<1, -1<\alpha<1$. Following distributional characteristics are presented in Chakraborty and Bhati (2016)
\begin{enumerate}
\item For $\alpha=0$, (\ref{e2}) reduces to $\mathcal{GD}(q)$ with pmf $p_y=(1-q)q^y, \quad y=0,1,\cdots, 0<q<1$.
\item For $\alpha=-1$, (\ref{e2}) reduces to a special case of the \textbf{Exponentiated Geometric distribution} of Chakraborty and Gupta (2015) with power parameter equal to 2. This is the \textbf{distribution of the maximum} of two iid $\mathcal{GD}(q)$ rvs.
\item For $\alpha=1$, (\ref{e2}) reduces to $\mathcal{GD}(q^2)$ with pmf $(1-q^2)q^{2y},$ which is the \textbf{distribution of the minimum} of two iid $\mathcal{GD}(q)$ rvs.
\item For $0<\alpha<1 (-1<\alpha<0)$ the $\mathcal{TGD}(q,\alpha)$ distribution with pmf given in (\ref{e2}), the ratio $p_y/p_{y-1}$,  $y=1,2,\cdots,$ forms a monotone increasing (decreasing) sequence.
\item $\mathcal{TGD}(q,\alpha)$ is unimodal with a nonzero mode for $-1<\alpha < -\left(q(2+q)\right)^{-1}$ provided $q>0.414$
\item The probability generating function(PGF) of $\mathcal{TGD}(q,\alpha)$ is given by \[
G_Y(z)=\frac{(1-q)(1-\alpha q(1-z)-q^2z)}{(1-qz)(1-q^2z)}, \qquad | q^2z|<1\]

\item The $r^{th}$ factorial moment of $Y \sim \mathcal{TGD}(q,\alpha)$ is given by
\[\mathbb{E}\left(Y_{(r)} \right)=(1-\alpha) r! \left(\frac{q}{1-q} \right)^r +\alpha r! \left(\frac{q^2}{1-q^2} \right)^r.\]
where $Y_{(r)}=Y(Y-1)...(Y-r+1).$
\end{enumerate}

\section{Reliability properties and Stochastic Ordering}
There are several situations in reliability where continuous time is not a good scale to measure the lifetime, in production we may interested in how many unit are produced by the machine before failure or health insurance companies are interested how long a patient stays in hospital before discharge/death. In such situations, the discrete hazard rate functions can be used to model ageing properties of discrete random lifetimes. We consider  different hazard rate function of $\mathcal{TGD}$ model and associated results as follows
\subsection{Reliability Properties}
\subsubsection{Hazard rate function and its classification}
The hazard rate function $r_X(x)$ for $X \sim \mathcal{TGD}(q,\alpha)$ is given as
\begin{align*}
r_X(x)=&\frac{P(X=x)}{{S_X(x)}}=\frac{(1-\alpha)q^x(1-q)+\alpha(1-q^2)q^{2x}}{(1-\alpha)q^x+\alpha q^{2x}} \\
=&\frac{(1-\alpha)(1-q)+\alpha q^x(1-q^2)}{(1-\alpha)+\alpha q^x}.
\end{align*}
\noindent The hazard rate function of $\mathcal{TGD}(q,\alpha)$ is plotted in Figure 1 for various values of parameters to investigate the monotonic properties and it is clear that the hazard rate of $\mathcal{TGD}(q,\alpha)$ is \textbf{increasing} for $-1<\alpha<0$, \textbf{decreasing} when $0<\alpha<1$ and \textbf{constant} if $\alpha=0$ or 1. Also it can be seen that even when $\alpha \neq 1$, the hazard rate approach to constant as $y$ increases. Smaller the value of $q$ the faster is the rate of stabilization of the hazard rate. \\

\begin{figure}
\begin{center}	
\includegraphics[scale=.65]{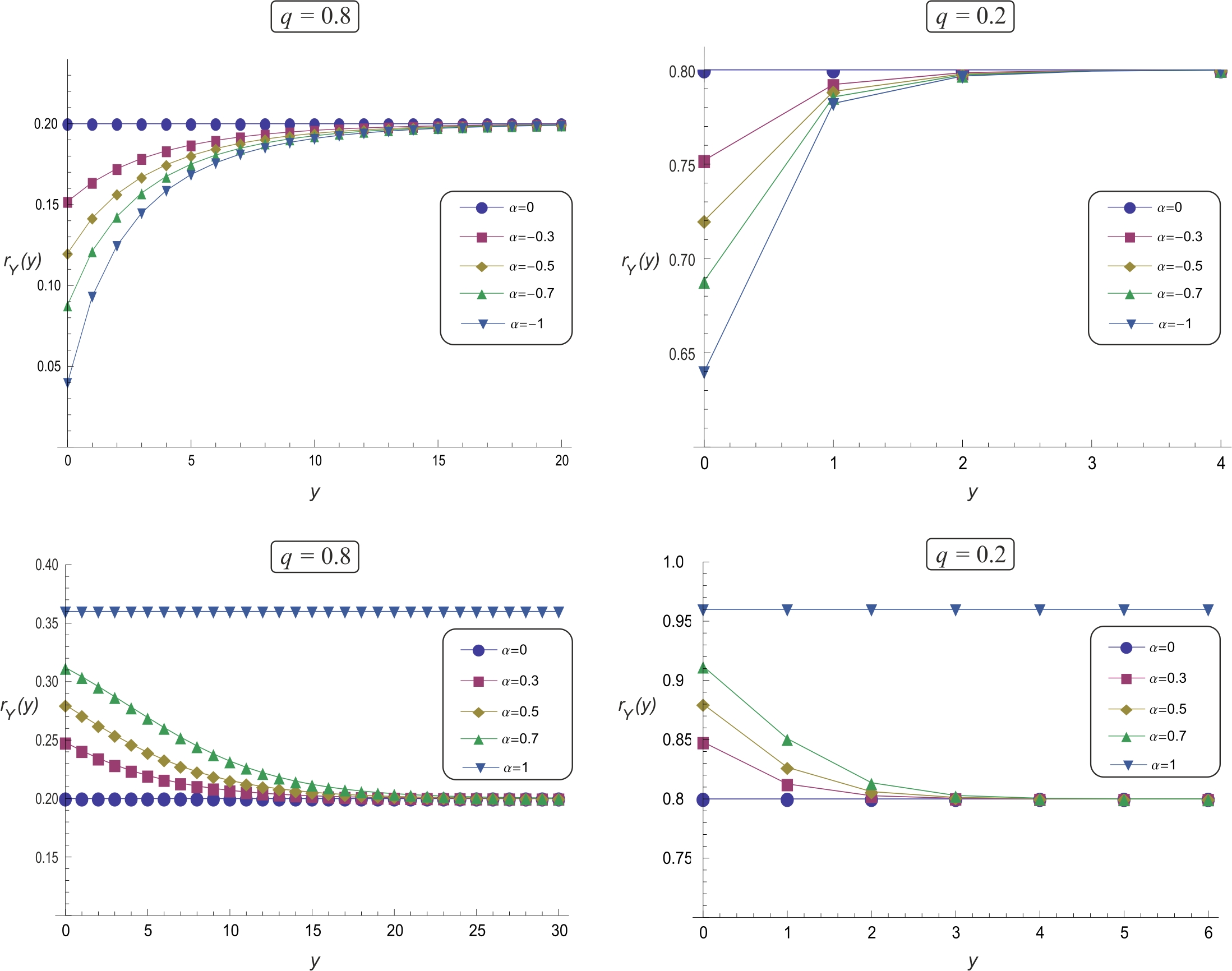}
\caption{Hazard rate function plots of $\mathcal{TGD}(q,\alpha)$.}
\end{center}
\end{figure}

\noindent \textbf{Theorem 1:} The $\mathcal{TGD}(q,\alpha)$ has increasing, decreasing and constant hazard rate for $-1<\alpha<0$, $0<\alpha<1$ and $\alpha=0$ or 1  respectively. \\
\textit{Proof:} The hazard rate of $\mathcal{TGD}(q,\alpha)$ is given as
\begin{eqnarray}
\nonumber r_Y(y)&=&\frac{(1-\alpha)(1-q)+\alpha q^y(1-q^2)}{(1-\alpha)+\alpha q^y}\\
\nonumber &=&1-q\frac{(1-\alpha)+\alpha q^{y+1}}{(1-\alpha)+\alpha q^y}.
\end{eqnarray}
But $q\frac{(1-\alpha)+\alpha q^{y+1}}{(1-\alpha)+\alpha q^y}$ is a decreasing(increasing) function of $y$ for $-1<\alpha<0(0<\alpha<1).$ Hence $r_Y(y)$ is increasing(decreasing)function of $y$ for $-1<\alpha<0(0<\alpha<1)$.
Constant hazard rates are obtained as $r_Y(y)=1-q$ for $\alpha=0$ and $r_Y(y)=1-q^2$ for $\alpha=1$. \\

\noindent \textbf{Remark} The hazard rate of $\mathcal{TGD}(q,\alpha)$ clearly obeys $r_Y(y)\le 1-q$ for $-1\le\alpha \le 0$ and $1-q \le r_Y(y)\le 1-q^2$ for $0 \le \alpha \le 1$.

\subsubsection{Second hazard rate}
The second rate of failure (Xie et al. (2002)) is given by
\[
r^*_Y(y)=\log\left(\frac{S_Y(y)}{S_Y(y+1)}\right)=\log\left(\frac{1-\alpha(1-q^y)}{q(1-\alpha(1-q^{y+1}))} \right)
\]

\subsubsection{Reversed hazard rate function}
\[
r^{**}_Y(y)=P(Y=y)/F_Y(y)=\frac{(1-\alpha)q^y(1-q)+\alpha q^{2y}(1-q^2)}{1-(1-\alpha)q^{y+1}-\alpha q^{2y+2}}
\]

\subsubsection{Mean residual life}
Kemp (2004) presented various characterization of discrete lifetime distribution among them the mean residual life(MRL) or life expectancy is an important characteristic, for $\mathcal{TGD}$, the closed expression for MRL is given as
\begin{equation} \label{e4}
L_Y(y)=\mathbb{E}\left(Y-y|Y\ge y\right)=\frac{1}{S_Y(y)}\sum\limits_{j>y}^{}S_Y(j)=\frac{q\left((1+q)(1-\alpha) +\alpha q^{y+1}\right)}{(1-q^2)(1-\alpha+\alpha q^y)}.
\end{equation}

\noindent \textbf{Theorem 2:} The mean residual life function   given in (\ref{e4}) is monotone decreasing (increasing) function of y depending on $-1<\alpha<0(0<\alpha<1).$ \\
\noindent \textit{Proof:} It can be easily be seen that
\[
\bigtriangleup L_Y(y)=L_Y(y+1)-L_Y(y)=\frac{(1-\alpha)\alpha q^{y+1}}{(1+q)\left(1-\alpha(1-q^y)\right)\left(1-\alpha(1-q^{y+1})\right)}.
\]
For any choice of $\alpha \in (-1,1)$ and $q \in (0,1)$, the denominator terms $\left( 1-\alpha(1-q^y)\right)$ and $\left( 1-\alpha(1-q^{y+1})\right)$ are always positive. Moreover, since $q \in (0,1)$, therefore $\bigtriangleup L_Y(y)<0$ for $-1<\alpha<0$ indicates decreasing mean residual life, whereas  $\bigtriangleup L_Y(y)>0$ for $0<\alpha<1$ indicates increasing mean residual life.

\subsection{Stochastic Ordering}
Many times there is a need of comparing the behaviour of one random variable with the other. Shaked and Shanthikumar (1994) has given many comparisons such as likelihood ratio order $(\preceq_{lr})$, the stochastic order $(\preceq_{st})$, the hazard rate order $(\preceq_{hr})$, the reversed hazard rate order $(\preceq_{rh})$ and the expectation order $(\preceq_{E})$ having various applications in different context. \\

\noindent \textbf{Theorem 3:} Let $Y$ be a random variable following $\mathcal{TGD}(q,\alpha)$ and $X$ be geometric random variable with parameter $p$. Then $R(z)=P(Y=z)/P(X=z)$ is an increasing(decreasing) function of $z$ for $-1<\alpha<0(0<\alpha<1)$ respectively i.e. $X\preceq_{lr}Y(X\succeq_{lr}Y)$. \\
\noindent \textit{Proof:} Since $R(z)=1+\alpha\left((1+q)q^z-1\right)$.
Thus, we have $R(z)\le(\ge)R(z+1)$ for $-1<\alpha<0(0<\alpha<1)$ for any $q \in (0,1)$. \\

\noindent \textbf{Corollary} Following results are direct implications of Theorem 3.

\begin{enumerate}
\item[i.] $X\preceq_{st}(\succeq_{st})Y$ that is, $P\left(X \ge z\right)\le(\ge)P\left(Y \ge z\right)$ for  $-1<\alpha<0(0<\alpha<1)$ respectively and for all $z$.
\item[ii.] $Y\preceq_{hr}(\succeq_{hr})X$ that is, $P\left(Y=z\right)/P\left(Y \ge z\right)\le(\ge)P\left(X=z\right)/P\left(X \ge z\right)$ for  $-1<\alpha<0(0<\alpha<1)$ respectively and for all $z$.
\item[iii.] $X\preceq_{rh}(\succeq_{rh})Y$ that is, $P\left(X=z\right)/P\left(X \le z\right)\le(\ge)P\left(Y=z\right)/P\left(Y \le z\right)$ for  $-1<\alpha<0(0<\alpha<1)$ respectively and for all $z$.
\item[iv.] $X\preceq_{E}(\succeq_{E})Y$ that is, $\mathbb{E}(X)\le(\ge)\mathbb{E}(Y)$ for  $-1<\alpha<0(0<\alpha<1)$ respectively and for all $z$.
\end{enumerate}

\noindent \textbf{Theorem 4:}
Let $Y_1$ and $Y_2$ be $\mathcal{TGD}(q_1,\alpha)$ and $\mathcal{TGD}(q_2,\alpha)$ respectively. Then $Y_2 \preceq_{st} Y_1$ iff $q_1\le q_2.$ \\
\noindent \textit{Proof:} We know that $Y_2\preceq_{st}Y_1$ iff $P(X_2 \ge y)\le P(Y_1\ge y)$ for all $y$, hence for $\mathcal{TGD}(q,\alpha)$ with $P(Y\ge y)=(1-\alpha)q^{2y}+\alpha q^y$ and it is clearly seen that
\[
(1-\alpha)q^{2y}_1+\alpha q^y_1 \le (1-\alpha)q^{2y}_2+\alpha q^y_2 \qquad \forall \, y  \qquad \text{iff} \qquad  q_1\le q_2.
\]
Hence $Y_2 \preceq_{st} Y_1$.

\section{Parameter Estimation and their \textbf{comparative} evaluation}
Estimates of the parameters $q$ and $\alpha$ of $\mathcal{TGD}$ model can be computed by following five methods (i) sample proportion of 1’s and 0’s method, (ii) sample quantiles, (iii) method of moments and finally (iv) maximum likelihood (ML) method and (v) ML via EM Algorithm. Moreover, in this section we carry out comparative study of ML estimator obtained numerically and via EM Algorithm utilizing initially estimate from one of the first three methods.

\subsection{\textbf{From sample proportion of 1's and 0's:}}
If $p_0, p_1$ be the known observed proportion of 0's and 1's in the sample, then the parameters $q$ and $\alpha$ can be estimated by solving the equations:
\[
p_0=(1-\alpha)(1-q)+\alpha(1-q^2) \quad  \text{and} \quad  p_1=(1-\alpha)q(1-q)+\alpha q^2(1-q^2)
\]

\subsection{From sample quantiles}
If $t_1, t_2$ be two observed points such that $F_Y(t_1)=\gamma_1, F_Y(t_2)=\gamma_2$, then  the two  parameters $q$ and $\alpha$ can be estimated by solving the simultaneous equations
\[
\gamma_1=1+(\alpha-1)q^{t_1+1}-\alpha q^{2(t_1+1)} \quad  \text{and} \quad  \gamma_2=1+(\alpha-1)q^{t_2+1}-\alpha q^{2(t_2+1)}.
\]

\subsection{Methods of Moments}
Denoting the first and second observed raw moments by $m_1$ and $m_2$ respectively, the moment estimates can be obtained by
\begin{itemize}
\item[a.] Either solving the following two equations simultaneously
\[
\frac{q(1-\alpha)+q^2}{1-q^2}=m_1 \quad  \text{and} \quad  \frac{q\left((1+q)^3-\alpha(q(3q+2)+1)\right)}{(1-q^2)^2}=m_2,
\]

\item[b.] or by the minimization method proposed by Khan et al. (1989) by minimizing $\left(\mathbb{E}(Y)-m_1\right)^2+\left(\mathbb{E}(Y^2)-m_2\right)^2$  with respect to $q$ and $\alpha$
\[
\left(\frac{q(1-\alpha)+q^2}{1-q^2}-m_1 \right)^2+\left(\frac{q\left((1+q)^3-\alpha(q(3q+2)+1)\right)}{(1-q^2)^2}-m_2 \right)^2
\]
\end{itemize}

\subsection{Maximum Likelihood Method}
Let $\mathbf{y}=(y_1, y_2,\cdots,y_n)^{\top}$ be a sample of $n$ observations drawn from $\mathcal{TGD}$ distribution, and $\Theta=\left(q,\alpha\right)^{\top}$ be the parametric vector. The $\log$-likelihood function for the corresponding sample is
\begin{equation}\label{e6}
l=\log L = n\log(1-q)+\log(q)\sum\limits_{i=1}^{n}y_i+\sum\limits_{i=1}^{n}\log\left((1-\alpha)+\alpha q^{y_i}(1+q)\right)
\end{equation}
and the score function $U(\Theta,\mathbf{y})=\left(\frac{\partial l_n}{\partial q},\frac{\partial l_n}{\partial \alpha}\right)^{\top}$ can be obtained by differentiating $\log$-likelihood function with respect to $q$ and $\alpha$ as
\begin{eqnarray}
\nonumber \frac{\partial l}{\partial q}&=& -\frac{n}{1-q}+\frac{1}{q}\sum\limits_{i=1}^{n}y_i+\sum\limits_{i=1}^{n}\frac{\alpha q^{y_i}+\alpha y_i(1+q)q^{y_i-1}}{1-\alpha+\alpha(1+q)q^{y_i}},\\
\nonumber \frac{\partial l}{\partial \alpha}&=& \sum\limits_{i=1}^{n}\frac{(1+q)q^{y_i}-1}{1-\alpha+\alpha(1+q)q^{y_i}}.
\end{eqnarray}
The maximum likelihood estimator(MLE) $(\hat{\Theta})$ of $\Theta$ is obtained by solving the non-linear system of equation $U(\Theta,\mathbf{y})=0$. Since the likelihood equations have no closed form solution, the estimator $\hat{q}$ and $\hat{\alpha}$ of the parameters $q$ and $\alpha$ can be obtained by maximizing $\log$-likelihood function using global numerical maximization techniques. Further, the Fisher’s information matrix is given by
\begin{equation} \label{e7}
\mathcal{I}_\mathbf{y}(q,\alpha)= \left(
\begin{array}{cc}
-\mathbb{E}\left(\frac{\partial ^2l}{\partial q^2}\right) & -\mathbb{E}\left(\frac{\partial ^2l}{\partial q\partial \alpha}\right) \\
-\mathbb{E}\left(\frac{\partial ^2l}{\partial q\partial \alpha}\right) & -\mathbb{E}\left(\frac{\partial ^2l}{\partial \alpha ^2}\right)
\end{array} \right) \approx \left(
\begin{array}{cc}
- \frac{\partial ^2l}{\partial q^2} & -\frac{\partial ^2l}{\partial q\partial \alpha } \\
- \frac{\partial ^2l}{\partial q\partial \alpha } & -\frac{\partial ^2l}{\partial \alpha ^2}
\end{array}
\right)_{q=\hat{q}, \alpha =\hat{\alpha}}
\end{equation}
\noindent where $\hat{q}$ and $\hat{\alpha}$ are the mle's of $q$ and $\alpha$ respectively, Moreover elements of $\mathcal{I}_\mathbf{y}(q,\alpha)$ are given as
\begin{align*}
\nonumber \frac{\partial^2 l}{\partial q^2}=&-\frac{n}{(1-q)^2}-\frac{1}{q^2}\sum\limits_{i=1}^{n}y_i-\sum_{i=1}^n \left(\frac{\alpha(1+q)(y_i-1)y_iq^{y_i-2}+2 \alpha y_i q^{y_i-1}}{1-\alpha +\alpha (1+q)q^{y_i}}\right.\\
\nonumber& \qquad \qquad \qquad \qquad \qquad \qquad  -\left.\left(\frac{\alpha (1+q)y_iq^{y_i-1}+\alpha q^{y_i}}{1-\alpha +\alpha (1+q)q^{y_i}}\right)^2\right), \\
\nonumber \frac{\partial^2 l}{\partial q\partial \alpha}=&\sum_{i=1}^n \left(\frac{(1+q)y_iq^{y_i-1}+q^{y_i}}{1-\alpha+\alpha(1+q)q^{y_i}}-\frac{\left(\alpha (1+q)y_iq^{y_i-1}+\alpha q^{y_i}\right)\left((1+q)q^{y_i}-1\right)}{\left(1-\alpha +\alpha (1+q)q^{y_i}\right)^2}\right), \\
\nonumber \frac{\partial^2 l}{\partial \alpha^2}=& -\sum_{i=1}^n \left(\frac{\left((1+q)q^{y_i}-1\right)^2}{1-\alpha +\alpha (1+q)q^{y_i}}\right).
\end{align*}

\subsection{MLE through EM Algorithm}
The Expected Maximization (EM) algorithm is an useful iterative procedure to compute ML estimators in the presence of missing data or assumed to have a missing values. The procedure follows with two steps called Expectation step(E-Step) and Maximization step(M-Step). The E-step concerns with the estimation of those data which are not observed whereas the M-step is a maximization step. for more details one may refer Dempster et al.(1977). \\
Let the complete-data be constituted with observed set of values $\mathbf{y}=(y_1,\cdots ,y_n)$ and the hypothetical data set $\mathbf{x}=(x_1,\cdots ,x_n)$, where the observations $y_i$'s are distributed with random variables $X$ defined as
\begin{equation}
X= \left\{
\begin{array}{ll}
1\quad  & w.p. \quad (1+\alpha)/2 \\
0\quad  & w.p. \quad (1-\alpha)/2
\end{array},
\right.
\end{equation}
and rv $Y$ be defined as
\begin{equation}
Y=XZ_{1:2}+(1-X)Z_{2:2},
\end{equation}%
where $Z_{1:2} \sim GD(q^2)$, $Z_{2:2} \sim \mathcal{EGD}(q,2)$(see Chakraborty and Gupta (2015)) and  $X_{i}\sim \mathit{Bernoulli}(\frac{1+\alpha}{2})$. \newline

\noindent Under the formulation, the \textbf{E-step} of an EM cycle requires
the expectation of $\left(X|Y;\Theta^{(k)}\right) $, where $\Theta
^{(k)}=(q^{(k)},\alpha^{(k)})$ is the current estimate
of $\Theta $ (in the $k^{th}$ iteration). Since the conditional distribution
of $X_{i}$ given $Y_{i}$ is
\begin{equation} \
\left(X_{i}|Y_{i},\Theta^{(k)}\right)\sim bernoulli\left( \frac{1+\alpha_{i}^{(k)}}{2}%
\right) ,
\end{equation}
with
\begin{equation}
\frac{1+\alpha_{i}^{(k)}}{2}= \frac{(1+\alpha^{(k)})\left(1-(q^{(k)})^2\right) (q^{(k)})^{2 y_i}}{(1+\alpha^{(k)}) \left(1-(q^{(k)})^2\right) (q^{(k)})^{2 y_i}+(1-\alpha^{(k)}) \left((1-q^{(k)}) (q^{(k)})^{y_i} \left(2-(q^{(k)}+1) (q^{(k)})^{y_i}\right)\right)}
\end{equation}%
where $\alpha^{(k)}$ is a set of known or estimated parameters at $k^{th}$
step with known initial values. Thus, by the property of the Binomial
distribution, the conditional mean is
\begin{equation}
\mathbb{E}(X_{i}|Y_{i},\Theta ^{(k)})=\left( \frac{1+\alpha_{i}^{(k)}}{2}\right) \quad and \quad \mathbb{V}(X_{i}|y)=\left( \frac{1+\alpha
_{i}^{(k)}}{2}\right) \left( \frac{1-\alpha_{i}^{(k)}}{2}\right) .
\label{e1}
\end{equation}%
For \textbf{M-step}: The likelihood function of joint pdf of hypothetical
complete-data $(Y_{i},X_{i}),i=1,\cdots ,n$ is given as
\begin{eqnarray*}
L^{\ast }(\Theta ;\mathbf{y,x})&=&\prod\limits_{i=1}^{n}\left( \frac{1+\alpha}{2}%
\right)^{x_{i}}  \left( \left(1-q^2\right) q^{2 y_i}\right)^{x_{i}} \\
&& \cdot \prod\limits_{i=1}^{n} \left( \frac{1-\alpha }{2}\right) ^{1-x_{i}} \left(\left((1-q) (q)^{y_i} \left(2-(1+q) q^{y_i}\right)\right)
\right) ^{1-x_{i}}
\end{eqnarray*}%
and the corresponding complete $\log $-likelihood function is given as
\begin{eqnarray} \nonumber
l_{n}^{\ast }(\Theta;\mathbf{x,y})&=& \log \left( \frac{1+\alpha}{2}\right)
\sum\limits_{i=1}^{n}x_{i}+\log \left( \frac{1-\alpha}{2}\right)
\sum\limits_{i=1}^{n}(1-x_{i}) + \log(1-q^2) \sum\limits_{i=1}^{n} x_i  \\ \nonumber
& &+ 2\log q \sum\limits_{i=1}^{n} x_i y_i+ \sum\limits_{i=1}^{n} (1-x_i) \left(y_i\log q +\log(1-q)+\log(2-q^{y_i}(1+q))\right) \\
\end{eqnarray}%
The components of the score function $U_{n}^{\ast }(\Theta )=\left( \frac{%
\partial l_{n}^{\ast }}{\partial \alpha},\frac{\partial l_{n}^{\ast}}{%
\partial q}\right)
^{\top }$ are given by
\begin{eqnarray}
\frac{\partial l_{n}^{\ast}}{\partial \alpha}= &&\frac{1}{1+\alpha}%
\sum\limits_{i=1}^{n}x_{i}-\frac{1}{1-\alpha}\sum%
\limits_{i=1}^{n}(1-x_{i}), \\ \nonumber
\frac{\partial l_{n}^{\ast }}{\partial q}= &-&\frac{2q}{1-q^2}\sum\limits_{i=1}^{n} x_i +\frac{2}{q}\sum\limits_{i=1}^{n}x_i y_i \\
& +&\sum\limits_{i=1}^{n} (1-x_i) \left(\frac{y_i}{q}-\frac{1}{1-q}-\frac{y_iq^{y_i-1}+(y_i+1)q^{y_i}}{2-q^{y_i}(1+q)} \right).
\end{eqnarray}%
The EM cycle will completed with the M-step by using the maximum likelihood
estimation over $\Theta$, i.e., $U_{n}^{\ast}(\mathbf{\widehat{\Theta};y,x})=0$ with the
unobserved $x_{i}s$ replaced by their conditional expectations given in (\ref%
{e1}). Hence we obtain the iterative procedure of the EM algorithm as
\begin{eqnarray*}
\widehat{\alpha}^{(k+1)}&=& \frac{1}{n}\sum\limits_{i=1}^{n}\alpha _{i}^{(k)},
\\
\widehat{q}^{(k+1)}&=& \frac{\sum\limits_{i=1}^{n}\left(\frac{1+\alpha_i^{(k)}}{2}\right) y_i}{\frac{2q^{(k+1)}}{1-(q^{(k+1)})^2}\sum\limits_{i=1}^{n} \left(\frac{1+\alpha_i^{(k)}}{2}\right)-\sum\limits_{i=1}^{n} \left(\frac{1-\alpha_i^{(k)}}{2}\right) \left(\frac{y_i}{q^{(k+1)}}-\frac{1}{1-q^{(k+1)}}-\frac{y_i(q^{(k+1)})^{y_i-1}+(y_i+1)(q^{(k+1)})^{y_i}}{2-(q^{(k+1)})^{y_i}(1+q^{(k+1)})} \right)},
\end{eqnarray*}%
\noindent where $\widehat{q}^{(k+1)}$ should be determined numerically.

\subsubsection{Standard errors of estimates obtained from EM-algorithm}

In this section, we obtain the standard errors (se) of the estimators from the
EM-algorithm using result of Louis (1982). Let $\mathbf{z}=(\mathbf{y},\mathbf{x})$, then the $2\times 2$
observed information matrix $I_{c}(\Theta,\mathbf{z})=\left[\frac{\partial
}{\partial\Theta}U_{c}(\Theta;\mathbf{z})\right] $ are given by
\begin{eqnarray*}
\frac{\partial^{2}l_{n}^{\ast }}{\partial \alpha ^{2}}&=& -\frac{1}{\left(
1+\alpha \right) ^{2}}\sum\limits_{i=1}^{n}x_{i}-\frac{1}{\left( 1-\alpha
\right)^{2}}\sum\limits_{i=1}^{n}(1-x_{i}),\\
\frac{\partial^{2}l_{n}^{\ast }}{\partial \alpha \partial q }&=&\frac{\partial ^{2}l_{n}^{\ast }}{\partial q \partial \alpha} =0, \\
\frac{\partial ^{2}l_{n}^{\ast }}{\partial q ^{2}}&=& -\frac{2\left(1+q^2\right)}{\left(1-q^2\right)^2} \sum\limits_{i=1}^{n}x_i -\frac{2}{q^2}\sum\limits_{i=1}^{n}x_i y_i -\sum\limits_{i=1}^{n} (1-x_i)\left(\frac{y_i}{q^2}+ \frac{q^{2y_i-2}(q y_i+q+y_i)^2}{\left(2-(q+1) q^{y_i}\right)^2} \right. \\  & &\left. + \frac{(y_i-1) y_i q^{y_i-2}+y_i (y_i+1) q^{y_i-1}}{2-(q+1) q^{y_i}}+\frac{1}{(1-q)^2} \right).
\end{eqnarray*}%
Taking the conditional expectation of $I_{c}(\Theta;\mathbf{z})$ given $x$, we
obtain the $2\times 2$ matrix
\begin{equation}
l_{c}(\Theta;\mathbf{y})=-\mathbb{E}(I_{c}(\Theta;z)|\mathbf{y})=(d_{ij}),
\end{equation}%
\noindent where
\begin{eqnarray*}
d_{11}&=&\frac{1}{\left( 1+\alpha \right) ^{2}}\sum\limits_{i=1}^{n}\mathbb{%
E}(X_{i}|y)+\frac{1}{\left( 1-\alpha \right)^{2}}\sum\limits_{i=1}^{n}(1-%
\mathbb{E}(X_{i}|\mathbf{y})), \\
d_{12}& =&d_{21}=0,\\
 d_{22}&=& \frac{2\left(1+q^2\right)}{\left(1-q^2\right)^2} \sum\limits_{i=1}^{n}\mathbb{E}(X_{i}|\mathbf{y}) +\frac{2}{q^2}\sum\limits_{i=1}^{n}\mathbb{E}(X_{i}|\mathbf{y}) y_i \\
 & &+\sum\limits_{i=1}^{n} (1-\mathbb{E}(X_{i}|\mathbf{y}))\left(\frac{y_i}{q^2}+ \frac{q^{2y_i-2}(q y_i+q+y_i)^2}{\left(2-(q+1) q^{y_i}\right)^2} \right. \\
 & &\left. + \frac{(y_i-1) y_i q^{y_i-2}+y_i (y_i+1) q^{y_i-1}}{2-(q+1) q^{y_i}}+\frac{1}{(1-q)^2} \right).
\end{eqnarray*}%
whereas computation of
\begin{equation}
l_{m}(\Theta;\mathbf{y})=\mathbb{V}\left(U_{c}(x;\theta)|\mathbf{y}\right) =m_{ij},
\end{equation}%
involve the following terms
\begin{eqnarray*}
&& m_{11}=\left( \frac{1}{1+\alpha }+\frac{1}{1-\alpha}\right)
^{2}\sum\limits_{i=1}^{n}\mathbb{V}(X_{i}|\mathbf{y}), \\
&& m_{12}=m_{21}= \sum\limits_{i=1}^{n} \left(\frac{1}{1+\alpha}+\frac{1}{1-\alpha}\right)\left(\frac{y_i}{q}-\frac{2q}{1-q^2}+\frac{1}{1-q}+\frac{y_i q^{y_i-1}+(y_i+1)q^{y_i}}{2-q^{y_i}(1+q)} \right) \mathbb{V}(X_{i}|\mathbf{y}),
\\
&& m_{22}=\sum\limits_{i=1}^{n}\left(\frac{y_i}{q}-\frac{2q}{1-q^2}+\frac{1}{1-q}+\frac{y_i q^{y_i-1}+(y_i+1)q^{y_i}}{2-q^{y_i}(1+q)} \right)^2 \mathbb{V}(X_{i}|\mathbf{y}). \\
\end{eqnarray*}%
Finally, the observed information matrix $(I)$ can be computed as
\[
I(\widehat{\Theta};\mathbf{y})=l_{c}(\widehat{\Theta};\mathbf{y})-l_{m}(\widehat{\Theta};\mathbf{y}),
\]
and $I(\widehat{\Theta};\mathbf{y})$ can be inverted to obtain an estimate of the
covariance matrix of the incomplete-data problem. The square roots of the
diagonal elements represent the estimates of the standard errors of the
parameters.

\subsection{Simulation Study \textbf{to evaluate EM algorithm}}
Here we study the behaviour of ML estimators obtained by direct numerical optimization and also through EM algorithm for different finite sample sizes and for different $\mathcal{TGD}(q,\alpha)$. Observations from $\mathcal{TGD}(q,\alpha)$  are generated using the quantile function provided in Chakraborty and Bhati (2016) (see result 4 of Table 1). In the next two subsections, first we investigate the performance of ML estimators $(\widehat{q},\widehat{\alpha})$ for various combinations of parameters $(q,\alpha)$ in subsection (3.6.1) and then evaluate the performance with respect to varying sample size for fixed parameter values in subsection (3.6.2).

\subsubsection{Performance of estimators for different parametric values}
A simulation study consisting of following steps is carried out for each triplet $(q,\alpha, n)$, considering $q = 0.25, 0.5, 0.75$, $\alpha = -0.70, -0.30, 0.30, 0.70$ and $n = 25, 50, 75, 100$.
\begin{enumerate}
\item Choose the value $(q_0,\alpha_0)$ for the corresponding elements of the parameter vector $\Theta= (q,\alpha)$, to specify the $\mathcal{TGD}(q,\alpha)$ ;
\item Choose sample size $n$;
\item Generate $N$ independent samples of size $n$ from $\mathcal{TGD}(q,\alpha)$;
\item Compute the ML and EM estimate $\widehat{\Theta}_n$ of $\Theta$  for each of the $N$ samples;
\item Compute the average bias, average standard error of the estimate.
\end{enumerate}
In our experiment we have considered the number of replication $N=1000$. It can be observed from Table 1 and Table 2 that as the sample size increase both average bias and average se both decreases.

\begin{table}[]
\small \addtolength{\tabcolsep}{-1.5pt}
\centering
\caption{Bias and MSE of Estimates computed by method of maximum likelihood and EM Algorithm method.}
\label{table1}
\begin{tabular}{cccccccc|cccc} \hline
\multicolumn{1}{l}{}    & \multicolumn{1}{l}{}   & \multicolumn{1}{l}{} & \multicolumn{1}{l}{} & \multicolumn{4}{c|}{MLE} & \multicolumn{4}{c}{EM Algorithm}        \\ \hline
\multicolumn{2}{c}{Parameters}    & $n$    &    & bias($\widehat{\alpha}$) & bias($\widehat{q}$) & se($\widehat{\alpha}$) & se($\widehat{q}$) & bias($\widehat{\alpha}$) & bias($\widehat{q}$) & se($\widehat{\alpha}$) & se($\widehat{q}$) \\ \hline
\multirow{4}{*}{$q$=0.25} & \multirow{4}{*}{$\alpha$= -0.75} & 25  &       & -0.5566 & 0.0099  & 1.5154     & 0.1144 & -0.0132 & 0.0232  & 0.9739     & 0.1147 \\
 &    & 50              &  & -0.2675   & 0.0101  & 0.9151   & 0.0866  & -0.0081  & 0.0122   & 0.7215     & 0.0835 \\
 &    & 75              &  & -0.1733   & 0.0050  & 0.6880   & 0.0694  & -0.0049  & 0.0073   & 0.5881     & 0.0677 \\
 &    & 100             &  & -0.1327   & 0.0053  & 0.5780   & 0.0600  & -0.0035  & 0.0052   & 0.5137     & 0.0589 \\
 &    &                 &  &   &       &         &          &         &          &          &                     \\ \hline
\multirow{4}{*}{$q$=0.5}  & \multirow{4}{*}{$\alpha$= -0.75} & 25              &                      & -0.1348 & -0.0149 & 0.5644     & 0.0859 & -0.0031 & -0.0058 & 0.5664     & 0.0854 \\
                        &                               & 50                   &                      & -0.0077 & -0.0001 & 0.3960     & 0.0619 & 0.0012  & -0.0029 & 0.3888     & 0.0601 \\
                        &                               & 75                   &                      & -0.0196 & -0.0012 & 0.3197     & 0.0498 & -0.0006 & -0.0011 & 0.3155     & 0.0489 \\
                        &                               & 100                  &                      & -0.0113 & -0.0026 & 0.2765     & 0.0432 & 0.0012  & -0.0028 & 0.2730     & 0.0424 \\
                        &                               &                      &                      &         &         &            &        &         &         &            &        \\ \hline
\multirow{4}{*}{$q$=0.75} & \multirow{4}{*}{$\alpha$= -0.75} & 25              &                      & -0.0411 & -0.0003 & 0.4012     & 0.0480 & 0.0060  & -0.0035 & 0.4190     & 0.0476 \\
                        &                               & 50                   &                      & -0.0085 & -0.0026 & 0.2766     & 0.0333 & -0.0002 & -0.0021 & 0.2964     & 0.0333 \\
                        &                               & 75                   &                      & -0.0011 & -0.0018 & 0.2242     & 0.0268 & 0.0012  & -0.0020 & 0.2337     & 0.0268 \\
                        &                               & 100                  &                      & -0.0008 & -0.0014 & 0.1909     & 0.0227 & -0.0009 & -0.0007 & 0.1990     & 0.0229 \\
                        &                               &                      &                      &         &         &            &        &         &         &            &        \\ \hline
\multirow{4}{*}{$q$=0.25} & \multirow{4}{*}{$\alpha$= -0.30} & 25                   &                      & -0.3455 & 0.0269  & 1.2484     & 0.1361 & -0.0340 & 0.0260  & 1.0043     & 0.1426 \\
                        &                               & 50                   &                      & -0.3055 & 0.0095  & 0.9006     & 0.1016 & -0.0240 & 0.0165  & 0.7203     & 0.1018 \\
                        &                               & 75                   &                      & -0.0391 & 0.0290  & 0.6466     & 0.0901 & -0.0125 & 0.0109  & 0.6045     & 0.0848 \\
                        &                               & 100                  &                      & -0.0997 & 0.0123  & 0.5770     & 0.0756 & -0.0090 & 0.0069  & 0.5269     & 0.0736 \\
                        &                               &                      &                      &         &         &            &        &         &         &            &        \\ \hline
\multirow{4}{*}{$q$=0.5}  & \multirow{4}{*}{$\alpha$= -0.30} & 25                   &                      & -0.0310 & 0.0045  & 0.6288     & 0.1097 & -0.0037 & 0.0000  & 0.6840     & 0.1153 \\
                        &                               & 50                   &                      & -0.0249 & 0.0011  & 0.4672     & 0.0803 & -0.0031 & -0.0009 & 0.4818     & 0.0818 \\
                        &                               & 75                   &                      & -0.0253 & -0.0002 & 0.3880     & 0.0668 & -0.0029 & -0.0008 & 0.3908     & 0.0664 \\
                        &                               & 100                  &                      & -0.0258 & -0.0008 & 0.3375     & 0.0580 & -0.0036 & -0.0004 & 0.3333     & 0.0568 \\
                        &                               &                      &                      &         &         &            &        &         &         &            &        \\ \hline
\multirow{4}{*}{$q$=0.75} & \multirow{4}{*}{$\alpha$= -0.30} & 25                   &                      & -0.0503 & 0.0000  & 0.5182     & 0.0625 & -0.0010 & -0.0046 & 0.6044     & 0.0689 \\
                        &                               & 50                   &                      & -0.0432 & 0.0010  & 0.3850     & 0.0459 & 0.0069  & -0.0030 & 0.4157     & 0.0482 \\
                        &                               & 75                   &                      & -0.0020 & 0.0003  & 0.3141     & 0.0369 & -0.0038 & 0.0000  & 0.3324     & 0.0381 \\
                        &                               & 100                  &                      & -0.0009 & 0.0000  & 0.2838     & 0.0330 & -0.0026 & -0.0005 & 0.2899     & 0.0332 \\ \hline
\end{tabular}
\end{table}

\begin{table}[]
\small \addtolength{\tabcolsep}{-1.5pt}
\centering
\caption{Bias and MSE of Estimates computed by method of maximum likelihood and EM Algorithm method.}
\label{table2}
\begin{tabular}{cccccccc|cccc} \hline
  &      &     &  & \multicolumn{4}{c}{MLE}     & \multicolumn{4}{c}{EM Algorithm}           \\ \hline
\multicolumn{2}{c}{Parameters}    & $n$    &    & bias($\widehat{\alpha}$) & bias($\widehat{q}$) & se($\widehat{\alpha}$) & se($\widehat{q}$) & bias($\widehat{\alpha}$) & bias($\widehat{q}$) & se($\widehat{\alpha}$) & se($\widehat{q}$) \\ \hline
\multirow{4}{*}{$q$=0.25} & \multirow{4}{*}{$\alpha$= 0.30} & 25  &  & -0.4174 & 0.0254 & 1.1108   & 0.1524 & -0.1138 & -0.0206 & 0.7619    & 0.1547 \\
&   & 50  &  & -0.2518 & 0.0178 & 0.8702    & 0.1281 & -0.0667 & -0.0158 & 0.8095    & 0.1632 \\
&   & 75  &  & -0.1338 & 0.0193 & 0.6331    & 0.1110 & -0.0481 & -0.0144 & 0.6889    & 0.1386 \\
&   & 100 &  & -0.0878 & 0.0215 & 0.5479    & 0.1013 & -0.0367 & -0.0032 & 0.6740    & 0.1353 \\
&   &     &  &          &         &            &         &          &          &            &         \\ \hline
\multirow{4}{*}{$q$=0.50} & \multirow{4}{*}{$\alpha$= 0.30} & 25  &  & -0.2343 & 0.0226 & 0.5962  & 0.1328 & -0.0404 & -0.0267 & 0.7990  & 0.1700 \\
&                      & 50  &  & -0.1440 & 0.0184 & 0.4884    & 0.1085 & -0.0335 & -0.0354 & 0.6296    & 0.1349 \\
&                      & 75  &  & -0.0611 & 0.0142 & 0.4132    & 0.0926 & -0.0319 & -0.0336 & 0.5801    & 0.1237 \\
&                      & 100 &  & -0.0586 & 0.0125 & 0.3970    & 0.0886 & -0.0213 & -0.0210 & 0.5013    & 0.1072 \\
&                      &     &  &          &         &            &         &          &          &            &         \\ \hline
\multirow{4}{*}{$q$=0.75} & \multirow{4}{*}{$\alpha$= 0.30} & 25  &  & -0.0594 & 0.0127 & 0.5713  & 0.0829 & -0.0143 & -0.0326 & 0.7882    & 0.1101 \\
&                      & 50  &  & -0.0316 & 0.0097 & 0.4540    & 0.0652 & -0.0173 & -0.0508 & 0.6689    & 0.0923 \\
&                      & 75  &  & -0.0250 & 0.0079 & 0.3969    & 0.0568 & -0.0177 & -0.0607 & 0.5449    & 0.0759 \\
&                      & 100 &  & -0.0081 & 0.0050 & 0.3729    & 0.0522 & -0.0107 & -0.0224 & 0.5029    & 0.0691 \\
&                      &     &  &          &         &            &         &          &          &            &         \\ \hline
\multirow{4}{*}{$q$=0.25} & \multirow{4}{*}{$\alpha$= 0.75} & 25  &  & -0.0975 & 0.0038 & 0.0240  & 0.0042 & -0.0234 & -0.0305 & 0.6862   & 0.1166 \\
 &                      & 50  &  & -0.0696 & 0.0138 & 0.0255    & 0.0027 & -0.0189 & -0.0220 & 0.5239    & 0.0423 \\
 &                      & 75  &  & -0.0995 & 0.0046 & 0.0751    & 0.0038 & -0.0125 & -0.0112 & 0.4443    & 0.0259 \\
 &                      & 100 &  & -0.0358 & 0.0070 & 0.0338    & 0.0027 & -0.0101 & -0.0071 & 0.4012    & 0.0125 \\
 &                      &     &  &          &         &            &         &          &          &            &         \\ \hline
\multirow{4}{*}{$q$=0.5}  & \multirow{4}{*}{$\alpha$= 0.75} & 25  &  & -0.1250 & 0.0288 & 0.5170   & 0.1474 & -0.0351 & -0.0112 & 0.5214  & 0.1627 \\
&   & 50  &  & -0.1162 & 0.0248 & 0.4238    & 0.1186 & -0.0158 & -0.0131 & 0.4862    & 0.1456 \\
&   & 75  &  & -0.0641 & 0.0140 & 0.3485    & 0.1000 & -0.0093 & -0.0583 & 0.3675    & 0.1088 \\
&   & 100 &  & -0.0493 & 0.0125 & 0.3422    & 0.0974 & -0.0037 & -0.1109 & 0.6810    & 0.1963 \\
&   &     &  &          &         &            &         &          &          &            &         \\ \hline
\multirow{4}{*}{$q$=0.75} & \multirow{4}{*}{$\alpha$= 0.75} & 25  &  & -0.1542 & 0.0350 & 0.5112    & 0.0966 & -0.0191 & -0.0176 & 0.5221   & 0.1048 \\
&   & 50  &  & -0.1114 & 0.0178 & 0.4014    & 0.0727 & -0.0350 & -0.0253 & 0.4722   & 0.0858 \\
&   & 75  &  & -0.0786 & 0.0100 & 0.3595    & 0.0638 & -0.1072 & -0.0986 & 0.3782   & 0.0676 \\
&   & 100 &  & -0.0455 & 0.0100 & 0.3139    & 0.0566 & -0.1168 & -0.1178 & 0.3662   & 0.0647 \\ \hline
\end{tabular}
\end{table}
\subsubsection{Performance of estimators for different sample size}
In this subsection, we assess the performance of ML estimators of $(q,\alpha)$ as sample size $n$, increases by considering $n = 25, 26, . . . ,200,$ for $q = 0.25$ and $\alpha = - 0.5$. For each $n$, we generate one thousand samples of size $n$ and obtain MLEs and their standard error. For each repetition we compute average bias and average squared error.

\begin{figure}
\begin{center}	
\includegraphics[scale=0.60]{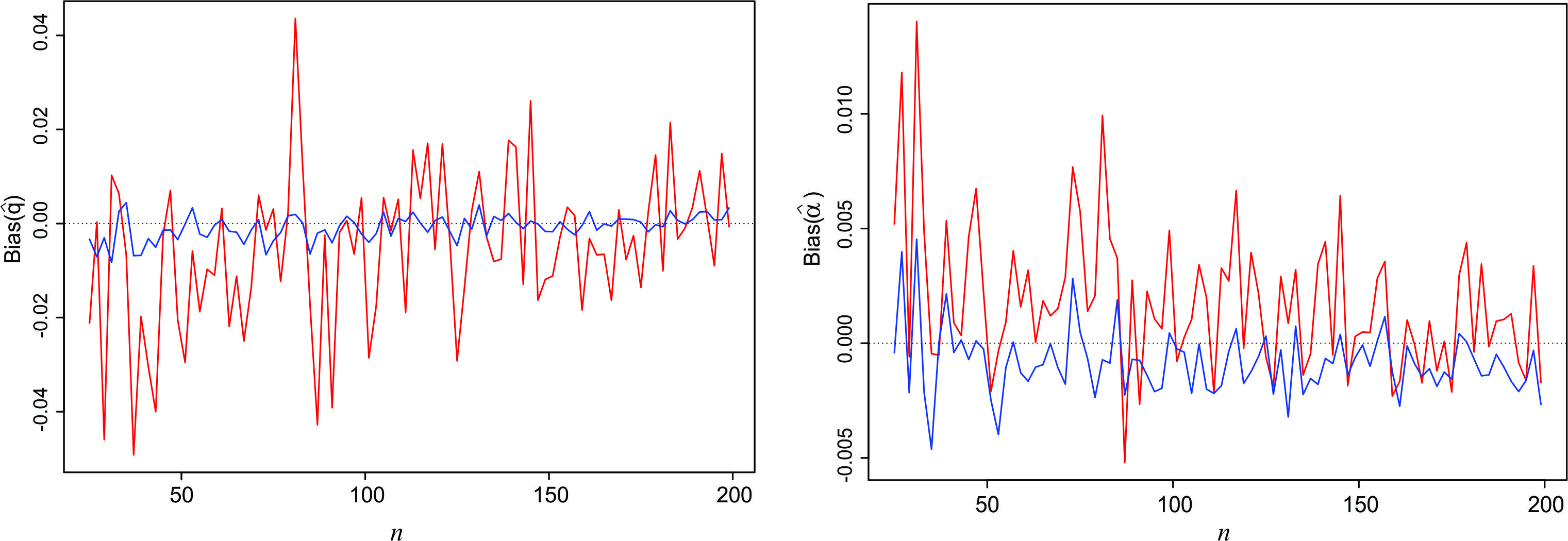}
\caption{Bias plot of estimated value of parameter $q$ and $\alpha$  for different sample sizes}
\end{center}
\end{figure}

\begin{figure}
\begin{center}	
\includegraphics[scale=0.60]{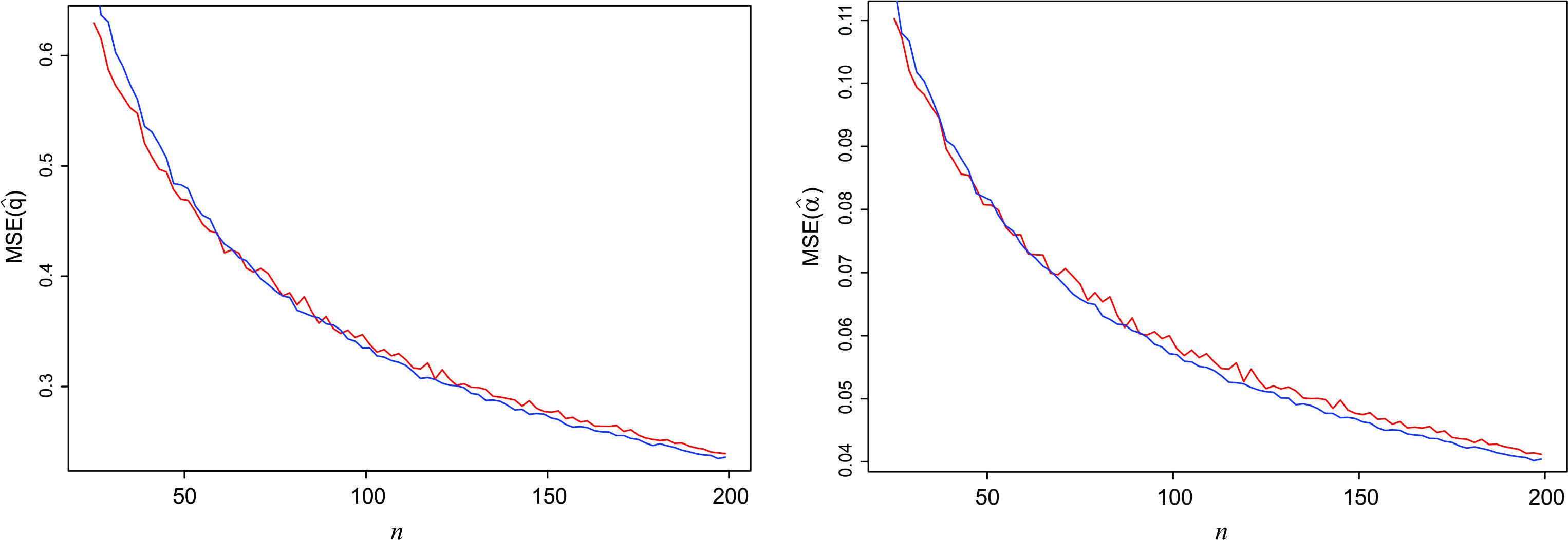}
\caption{MSE plot of estimated value of parameter $q$ and $\alpha$  for different sample sizes}
\end{center}
\end{figure}
Figures 2 and 3 shows behaviour of average bias and average standard error of parameter $q$ and $\alpha$, for fixed $q = 0.25$ and $\alpha = -0.5$, as one varies sample size $n$. The horizontal dotted lines in Figure 2 corresponds to zero value and it is clear in figure 2 that the biases approach to zero with increasing $n$ also in figure 3, average standard errors for both parameters ($q$ and $\alpha$) decrease with increase in $n$. Similar observations were also noted for other parametric values.

Based on our findings it is clear that EM algorithm produces better ML estimators with smaller average bias as compared to the regular ML estimators while w.r.t. standard error there is not much to choose between the two procedures.

\section{\textbf{Tests of hypothesis}}
The $\mathcal{TGD}(q,\alpha)$ distribution with parameter vector  $\Theta=\left(q,\alpha\right)^{\top}$ reduces to the Geometric distribution with
parameter $q$ when $\alpha=0$. This additional parameter $\alpha$ controls the proportion of zeros of the distribution relative to geometric distribution and also the tail length. Therefore it is of interst to develop test procedure for detecting departure of $\alpha$ from $0$. In this section we develop the likelihood ratio test (LRT), the Rao's score test and  the Wald's test for testing the null hypothesis $\mathcal{H}_{0}:\alpha=0$ against the alternative hypothesis $\mathcal{H}_{1}$ : $\alpha\ne 0$ and numerically study the statistical power of these tests  through extensive simulation.

\subsection{\textbf{Likelihood Ratio Test, Rao's Score Test and Wald's Test}}

The Likelihood Ratio Test(LRT) is based on the difference between the maximum of the likelihood under null and the alternative hypotheses. The LRT test statistics is given by $-2 \log(\frac{L(\widehat{\Theta}^{*};\mathbf{y})}{L(\widehat{\Theta};\mathbf{y})})$ where $\widehat{\Theta}^{*}$ and $\widehat{\Theta}$ are the MLE obtained under the null and alternative hypotheses respectively. The LRT is generally employed to test the significance of the additional parameter which is included to extend a base model.

The Rao's Score test (Rao, 1948)is based on the score vector defined as the first derivative of the log likelihood function w.r.t. the parameters. Rao's score test statistic $U I^{-1}U^{/}$, where $U$ is the score vector and $I$ is the information matrix derived under the null hypothesis. The score vector and the information matrix, obtained by evaluating the derivative of the log-likelihood function, $\log L$  are provided in section $4.4$.Note that the scores actully are the slopes of the likelihood functions.

The Wald's test statistics (1943)is based on on the difference between the  maximum of the likelihood estimate  value of the parameter under alternative hypothesis and the value specified by the under null hypothesis. The Wald's test statistic is given in our case by $(\widehat{\alpha}-\alpha_{0}) I_{[22]}^{-1}(\widehat{\alpha}-\alpha_{0})^{/}$, where  $I_{[22]}^{-1}$  is the $(2,2)th$ element of the inverse of the information matrix $I$, and $\widehat{\alpha}$ is the MLE of $\alpha$ both under alternative hypotheses. Whereas $\alpha_{0}$ is the value of $\alpha$ as per $H_{0}$. Note that $I_{[22]}^{-1}$ is an estimate of the variance of $\alpha$. Therefore in the present case our Wald's statistic reduces to $(\widehat{\alpha})^{2} \mathbb{V}(\widehat{\alpha})$.

All the test statistics follow asymptotically Chisqure distribution with ``$k$'' degrees of freedom, where ``$k$'' is the number of parameter specified by the null hypothesis. so in the present case the df is just ``$1$''. For well behaved likelihood function all these tests are based on measuring the discrepancy between null and the alternative hypotheses.

\subsection{\textbf{Statistical Power Analysis}}
Here we present a simulation based study of the statistical power of LR tests, Rao's Score test and the Wald's test considering $5\%$level of significance.Since the test are asymptotic in nature  we have considered four different sample sizes, two samples of smaller sizes namely $ n=100, 300$, one medium size $500$ and one large size $1000$.We have generated $1000$ replications for each sample size $n$. The power of these test are estimated by proportion of rejection in these $1000$ replications. The effect size (ES) is a measure of departure from the null hypothesis which in the present case is given by $\alpha-0=\alpha$ is fixed at $-0.7,0.5,-0.3,-0.1,0.1,0.2,0.5,0.7$ for our experiments.

The results are presented in Table 3, Table 4, Figures 4 to 7 reveal that the as expected the power increases with the sample size $n$ and ES; for positive ES all the tests displays show increase in power with the increase in either or both ES and sample size, while for negative ES power increases in a much faster pace. Power for score test is more than LRT for negative effect size where as it is other way for positive effect size. For positive effect size the power of the tests gets closer with increase in sample size.From the over all observation it is clear that the Wald,s test is more reliable than both LRT and Score tests.

\begin{table}[]
\small \addtolength{\tabcolsep}{-2pt}
\centering
\caption{}
\begin{tabular}{r|ccc|ccc|ccc|ccc}
\multicolumn{1}{l}{}    & \multicolumn{12}{c}{$q$=0.30}     \\ \hline
\multicolumn{1}{c|}{n}   & \multicolumn{3}{c|}{100}    & \multicolumn{3}{c|}{300}     & \multicolumn{3}{c|}{500}       & \multicolumn{3}{c}{1000}   \\ \hline
\multicolumn{1}{c|}{$\alpha$}   & LR     & Score   & Wald    & LR     & Score   & Wald     & LR     & Score   & Wald       & LR     & Score  & Wald     \\ \hline
-0.7 & 0.305 & 0.565 & 0.127 & 0.742 & 0.851 & 0.741 & 0.922 & 0.963 & 0.927 & 0.998 & 0.999 & 0.999 \\
-0.5 & 0.137 & 0.303 & 0.047 & 0.412 & 0.537 & 0.389 & 0.619 & 0.707 & 0.620 & 0.917 & 0.942 & 0.924 \\
-0.3 & 0.074 & 0.177 & 0.047 & 0.147 & 0.219 & 0.129 & 0.207 & 0.274 & 0.200 & 0.457 & 0.519 & 0.462 \\
-0.1 & 0.049 & 0.098 & 0.064 & 0.059 & 0.076 & 0.065 & 0.065 & 0.076 & 0.053 & 0.089 & 0.109 & 0.085 \\
0.1  & 0.041 & 0.072 & 0.086 & 0.052 & 0.055 & 0.101 & 0.056 & 0.054 & 0.060 & 0.080 & 0.071 & 0.051 \\
0.3  & 0.034 & 0.090 & 0.129 & 0.082 & 0.101 & 0.151 & 0.153 & 0.156 & 0.180 & 0.296 & 0.292 & 0.202 \\
0.5  & 0.043 & 0.153 & 0.172 & 0.139 & 0.213 & 0.271 & 0.289 & 0.336 & 0.347 & 0.546 & 0.575 & 0.455 \\
0.7  & 0.276 & 0.181 & 0.468 & 0.265 & 0.300 & 0.555 & 0.367 & 0.424 & 0.627 & 0.634 & 0.642 & 0.725 \\ \hline
\multicolumn{13}{c}{}  \\
\multicolumn{1}{l}{}      & \multicolumn{12}{c}{$q$=0.45} \\ \hline
\multicolumn{1}{c|}{n}     & \multicolumn{3}{c|}{100}  & \multicolumn{3}{c|}{300}    & \multicolumn{3}{c|}{500}   & \multicolumn{3}{c}{1000}  \\ \hline
\multicolumn{1}{c|}{$\alpha$} & LR   & Score  & Wald  & LR  & Score  & Wald  & LR   & Score   & Wald   & LR   & Score  & Wald              \\ \hline
-0.7 & 0.470 & 0.787 & 0.563 & 0.933 & 0.982 & 0.956 & 0.989 & 0.997 & 0.993 & 1.000 & 1.000 & 1.000 \\
-0.5 & 0.241 & 0.540 & 0.310 & 0.611 & 0.792 & 0.675 & 0.835 & 0.909 & 0.873 & 0.993 & 0.996 & 0.994 \\
-0.3 & 0.089 & 0.279 & 0.125 & 0.223 & 0.368 & 0.280 & 0.325 & 0.465 & 0.391 & 0.641 & 0.729 & 0.699 \\
-0.1 & 0.058 & 0.157 & 0.089 & 0.076 & 0.128 & 0.105 & 0.071 & 0.106 & 0.085 & 0.090 & 0.137 & 0.122 \\
0.1  & 0.035 & 0.083 & 0.078 & 0.060 & 0.062 & 0.096 & 0.059 & 0.055 & 0.071 & 0.097 & 0.073 & 0.051 \\
0.3  & 0.033 & 0.062 & 0.117 & 0.117 & 0.083 & 0.171 & 0.210 & 0.163 & 0.193 & 0.396 & 0.313 & 0.233 \\
0.5  & 0.055 & 0.106 & 0.199 & 0.224 & 0.200 & 0.316 & 0.417 & 0.351 & 0.427 & 0.700 & 0.645 & 0.539 \\
0.7  & 0.268 & 0.121 & 0.468 & 0.347 & 0.227 & 0.639 & 0.497 & 0.377 & 0.711 & 0.763 & 0.685 & 0.805 \\ \hline
\end{tabular}
\end{table}

\begin{table}[]
\small \addtolength{\tabcolsep}{-2pt}
\centering
\caption{}
\label{my-label}
\begin{tabular}{r|ccc|ccc|ccc|ccc}
\multicolumn{1}{l}{}    & \multicolumn{12}{c}{$q$=0.6}     \\ \hline
\multicolumn{1}{c|}{n}   & \multicolumn{3}{c|}{100}    & \multicolumn{3}{c|}{300}     & \multicolumn{3}{c|}{500}       & \multicolumn{3}{c}{1000}   \\ \hline
\multicolumn{1}{c|}{$\alpha$}   & LR     & Score   & Wald    & LR     & Score   & Wald     & LR     & Score   & Wald       & LR     & Score  & Wald     \\ \hline
-0.7 & 0.628 & 0.888 & 0.760 & 0.985 & 0.997 & 0.996 & 1.000 & 1.000 & 1.000 & 1.000 & 1.000 & 1.000 \\
-0.5 & 0.281 & 0.630 & 0.434 & 0.745 & 0.875 & 0.825 & 0.920 & 0.962 & 0.947 & 0.997 & 0.998 & 0.998 \\
-0.3 & 0.120 & 0.351 & 0.213 & 0.273 & 0.457 & 0.364 & 0.453 & 0.627 & 0.550 & 0.748 & 0.830 & 0.802 \\
-0.1 & 0.045 & 0.178 & 0.113 & 0.070 & 0.139 & 0.110 & 0.081 & 0.125 & 0.108 & 0.109 & 0.180 & 0.150 \\
0.1  & 0.042 & 0.103 & 0.108 & 0.046 & 0.054 & 0.097 & 0.072 & 0.046 & 0.078 & 0.113 & 0.080 & 0.057 \\
0.3  & 0.043 & 0.077 & 0.141 & 0.143 & 0.082 & 0.195 & 0.267 & 0.165 & 0.208 & 0.450 & 0.358 & 0.257 \\
0.5  & 0.064 & 0.089 & 0.202 & 0.252 & 0.172 & 0.350 & 0.481 & 0.336 & 0.415 & 0.784 & 0.689 & 0.583 \\
0.7  & 0.265 & 0.083 & 0.485 & 0.392 & 0.188 & 0.667 & 0.563 & 0.387 & 0.741 & 0.817 & 0.697 & 0.863 \\ \hline
\multicolumn{13}{c}{}  \\
\multicolumn{1}{l}{}      & \multicolumn{12}{c}{$q$=0.75} \\ \hline
\multicolumn{1}{c|}{n}     & \multicolumn{3}{c|}{100}  & \multicolumn{3}{c|}{300}    & \multicolumn{3}{c|}{500}   & \multicolumn{3}{c}{1000}  \\ \hline
\multicolumn{1}{c|}{$\alpha$} & LR   & Score  & Wald  & LR  & Score  & Wald  & LR   & Score   & Wald   & LR   & Score  & Wald              \\ \hline
-0.7 & 0.698 & 0.941 & 0.852 & 0.994 & 1.000 & 0.998 & 1.000 & 1.000 & 1.000 & 1.000 & 1.000 & 1.000 \\
-0.5 & 0.336 & 0.686 & 0.532 & 0.798 & 0.921 & 0.868 & 0.956 & 0.987 & 0.979 & 0.998 & 0.999 & 0.999 \\
-0.3 & 0.142 & 0.374 & 0.245 & 0.297 & 0.512 & 0.418 & 0.488 & 0.669 & 0.600 & 0.810 & 0.877 & 0.860 \\
-0.1 & 0.045 & 0.188 & 0.131 & 0.057 & 0.150 & 0.119 & 0.077 & 0.153 & 0.130 & 0.090 & 0.167 & 0.141 \\
0.1  & 0.045 & 0.112 & 0.115 & 0.060 & 0.056 & 0.110 & 0.095 & 0.071 & 0.104 & 0.092 & 0.057 & 0.045 \\
0.3  & 0.044 & 0.072 & 0.139 & 0.145 & 0.076 & 0.179 & 0.299 & 0.182 & 0.236 & 0.470 & 0.356 & 0.245 \\
0.5  & 0.091 & 0.078 & 0.242 & 0.285 & 0.164 & 0.355 & 0.491 & 0.321 & 0.427 & 0.789 & 0.683 & 0.586 \\
0.7  & 0.316 & 0.089 & 0.525 & 0.412 & 0.176 & 0.657 & 0.585 & 0.369 & 0.764 & 0.845 & 0.698 & 0.860 \\ \hline
\end{tabular}
\end{table}

\begin{figure}
\begin{center}	
\includegraphics[scale=0.60]{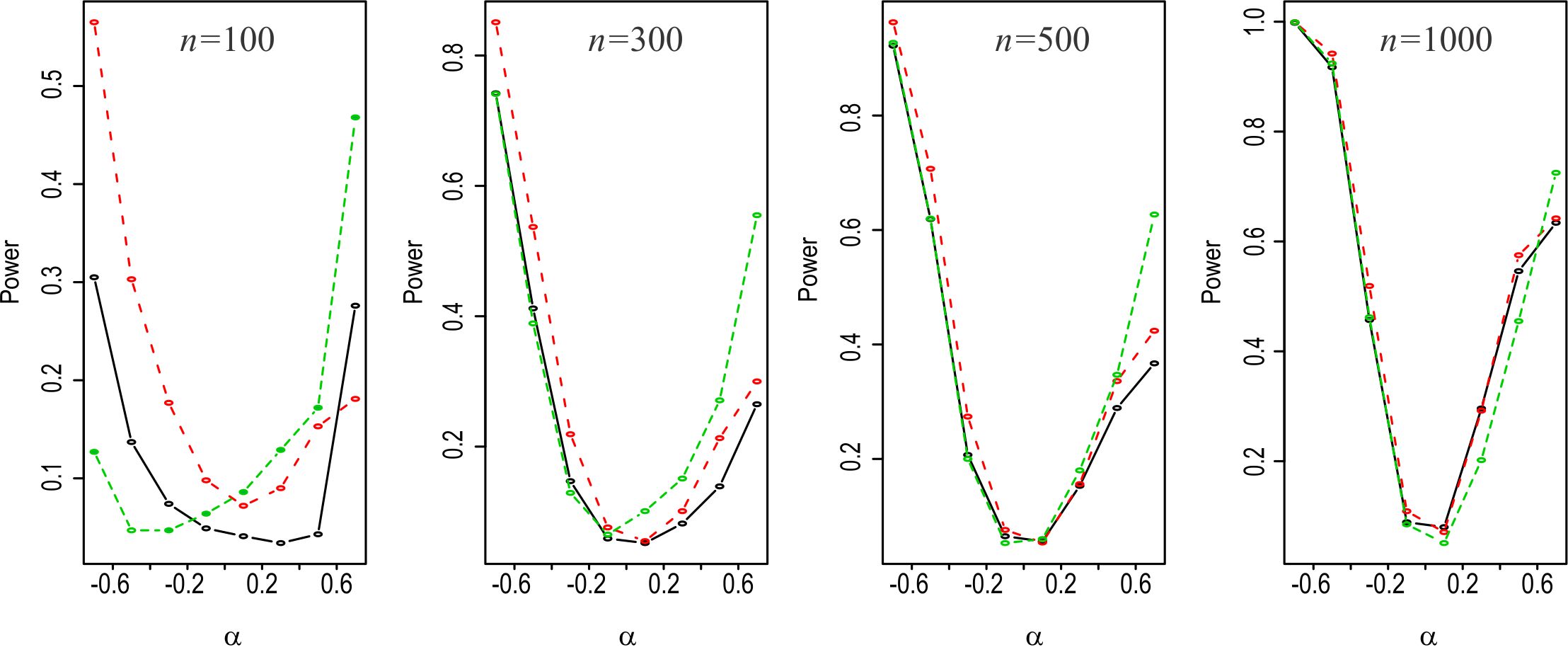}
\caption{Power curve of LRT(black), Score Test(Red) and Wald's Test(Green) for different $n$ and $q=0.3$.}
\end{center}
\end{figure}
\begin{figure}
\begin{center}	
\includegraphics[scale=0.60]{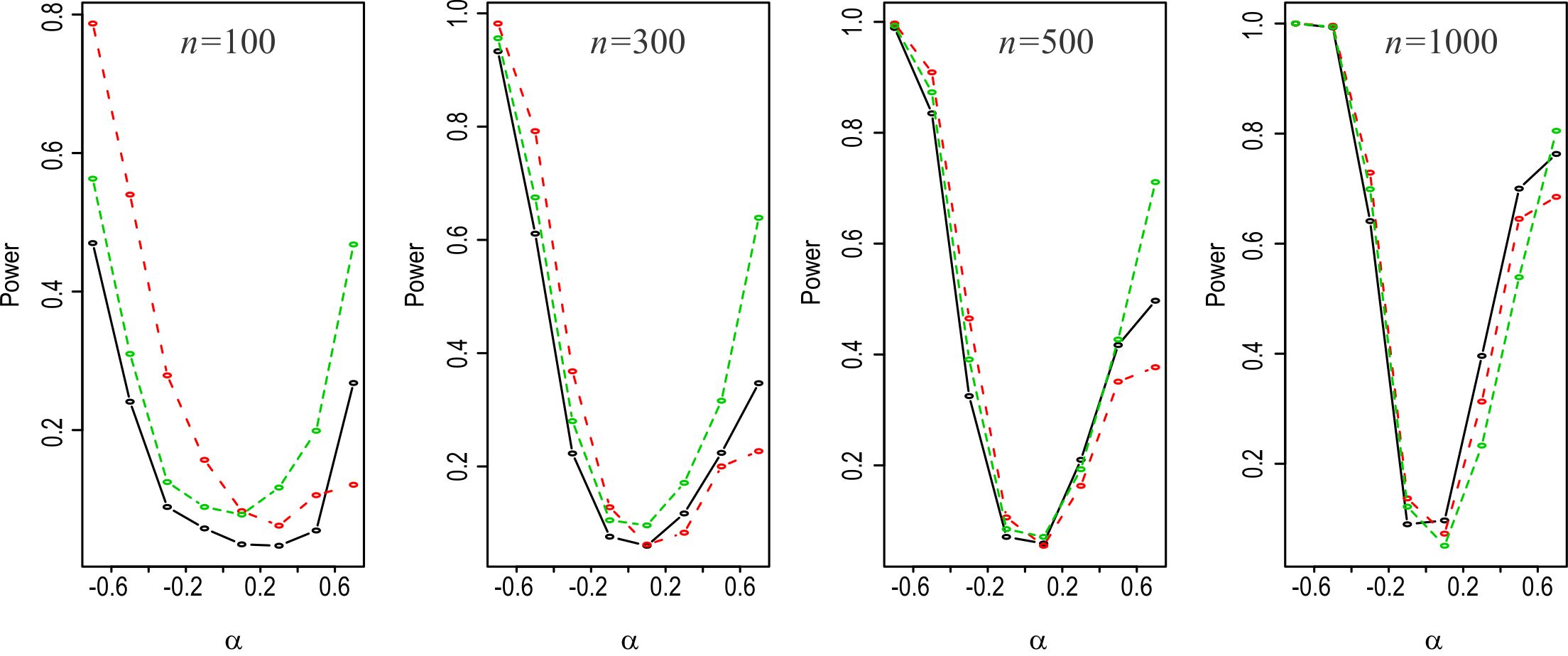}
\caption{Power of LRT(black), Score Test(Red) and Wald's Test(Green) for different $n$ and $q=0.45$.}
\end{center}
\end{figure}
\begin{figure}
\begin{center}	
\includegraphics[scale=0.60]{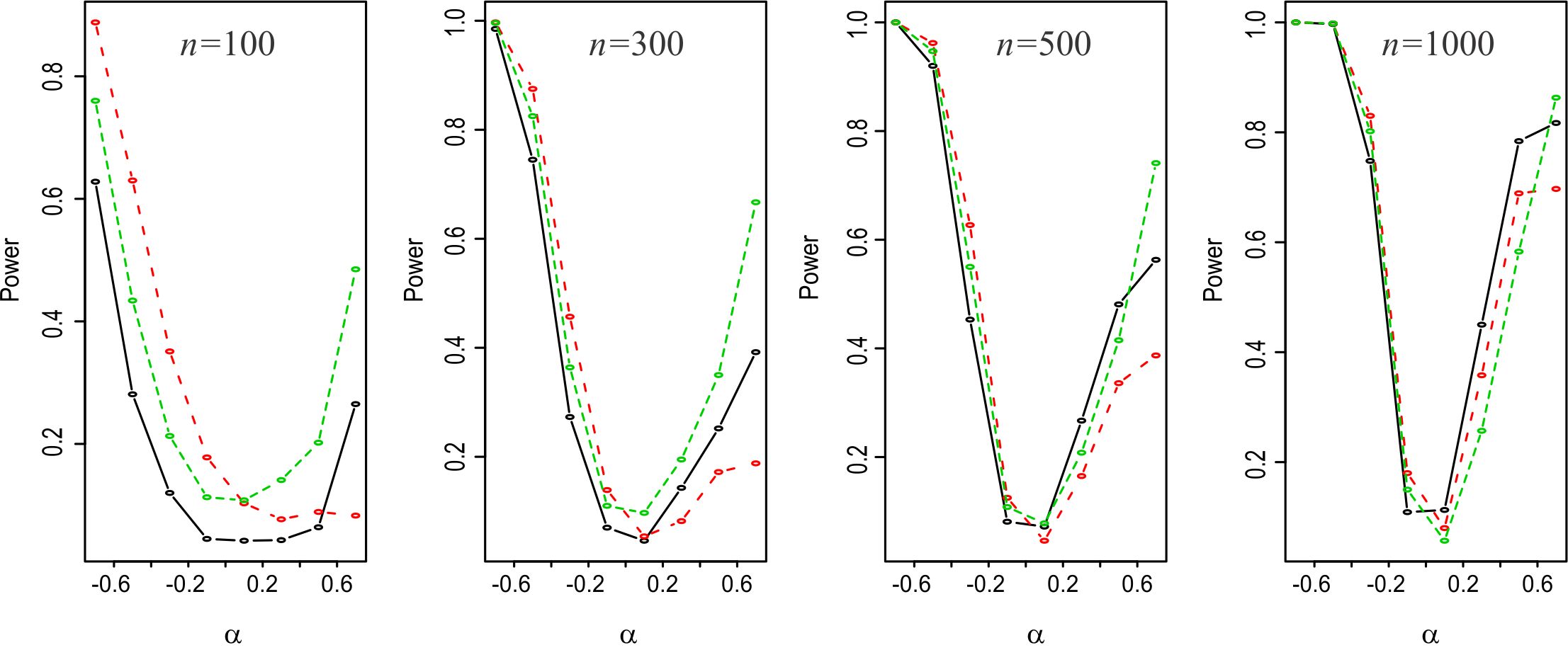}
\caption{Power of LRT(black), Score Test(Red) and Wald's Test(Green) for different $n$ and $q=0.6$. }
\end{center}
\end{figure}
\begin{figure}
\begin{center}	
\includegraphics[scale=0.60]{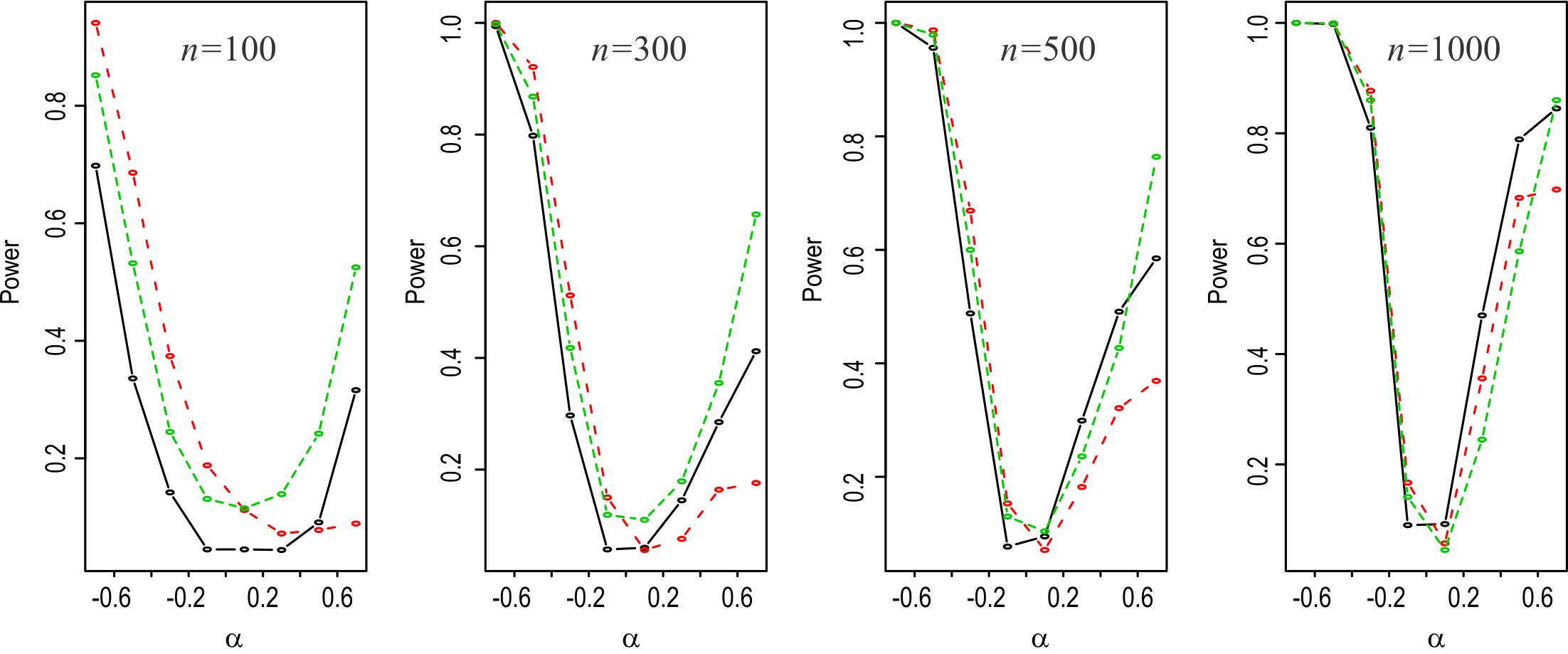}
\caption{Power of LRT(black), Score Test(Red) and Wald's Test(Green) for different $n$ and $q=0.75$. }
\end{center}
\end{figure}

\section{Data Analysis}
For the purpose of illustration, in this section, we consider following two data sets with details as follows:

\begin{enumerate}

\item[i.] \textit{Number of Fires in Greece} (\texttt{NTG}) \\
The data comprise of numbers of fires in district forest of Greece from period 1 July 1998 to 31 August 1998. The observed sample values pf size 123 for these data are the following(frequency in parentheses and none when it is equal to one):
0(16),1(13), 2(14),  3(9),  4(11),  5(13),  6(8),  7(4),  8(9),  9(6), 10(3), 11(4), 12(6), 15(4), 16, 20, 43. The data were previously studied by Bakourch et al. (2014) and Karlis and Xekalaki (2001).

\item[ii.] \textit{Number of doctor visits} (\texttt{Doctor\_Visit}) \\
This data is about the number of doctor consultations in a two-week period from the 1977-78 Australian Health Surveys (see Cameron and Trivedi (1998)) and is as follows: 0(4141), 1(782), 2(174), 3(30), 4(24), 5(39).
\end{enumerate}

The null hypothesis $\mathcal{H}_0:\alpha = 0$ against $\mathcal{H}_1:\alpha \ne 0$ are examined utilizing the LR, Rao's Score and Wald's test, and the results along with the descriptive statistics are presented in Table 5. Both the datasets confirm the presence of over dispersion. Moreover Rao's Score and Wald's test rejects the null hypothesis at 5\% significance level. The suitability of the proposed $\mathcal{TGD}(q,\alpha)$ model with other competitive distributions namely Com-Poisson $(p,\alpha)$ (Conway and Maxwell (1962)), $\mathcal{ZDGGD}(q,\alpha)$ (Sastry et al. (2016)), Negative Binomial$(r,p)$ is carried out and the log likelihood and Akaiki Information Criteria(AIC) value are computed for four models for both the datasets. The results in table 6 reveals that the $\mathcal{TGD}(q,\alpha)$ is the best fitted model and could be consider as competitive model for the datasets considered.

\begin{table}[h]
\centering
\caption{Descriptive and Test Statistic for both the datasets}
\label{my-label}
\begin{tabular}{lrrrrrr} \hline
Data set & Mean  & Variance & Index of dispersion & LRT & Score Test & Wald's Test  \\ \hline
\texttt{NTG} &  5.398 & 30.045     & 5.565  & 3.567	& 41.018	& 5.445\\
\texttt{Doctor\_Visit} &  0.291 &  0.514     & 1.765 & 96.34	& 116.33 & 247.321  \\ \hline
\end{tabular}
\end{table}

\begin{table}[h]
\centering
\caption{Comparative study of data fitting}
\label{my-label}
\begin{tabular}{llrrrr} \hline
 &   & \multicolumn{1}{c}{$\mathcal{NB}(r,p)$} & \multicolumn{1}{c}{Com-Pois$(p,\alpha)$} & \multicolumn{1}{c}{$\mathcal{ZDGGD}(q,\alpha)$} & \multicolumn{1}{c}{$\mathcal{TGD}(q,\alpha)$} \\ \hline
\multirow{3}{*}{\texttt{NTG}} & MLE & (1.336,0.802) & (0.947,0.055)   & (0.838,-0.207)    & (0.811, -0.465)              \\
                            & LL           & -339.649      & -339.843        & -340.742          & -339.354                      \\
                            & AIC          & 683.299       & 683.686         & 685.485           & 682.708                       \\ \hline
\multirow{3}{*}{\texttt{Doctor\_Visit}} & MLE &(0.439, 0.399)  & (0.225, -3.612)  & (0.3057, 0.3493)   & (0.386, 0.755)                \\
                            & LL           & -3533.28      & -3576.78        & -3542.53          & -3528.61                      \\
                            & AIC          & 7070.56       & 7157.55         & 7089.07           & 7061.21                       \\ \hline
\end{tabular}
\end{table}

\section{Conclusion}

The current paper investigates some additional property of the $\mathcal{TGD}(q,\alpha)$ distribution with emphasis on the simulation study of the behaviors of the parameter estimation and also power of tests of hypothesis to check statistical significance of the additional parameter. In the parameter estimation we have  presented different methods including the EM algorithm implementation of the MLE. A comparative simulation based evaluation of the EM algorithm based MLE against the usual MLE has reveled the superiority of the former in terms of the bias and mean squared errors. We have also presented data modeling examples to showcase the advantage of the $\mathcal{TGD}(q,\alpha)$ over some of the existing distribution from literature. As such it is envisaged that the present contribution will useful for discrete data analysts.

\end{document}